\documentclass{article}
\usepackage{template/spconf}
\usepackage{amsmath}
\usepackage{graphicx}
\usepackage{booktabs}
\usepackage{threeparttable}
\usepackage{appendix}
\usepackage{algorithm}
\usepackage{algorithmic}
\usepackage{makecell}
\usepackage{multirow}
\usepackage{hyperref}
\usepackage{amsfonts}
\usepackage{bm}

\newcommand{\figref}{Fig.~\ref}
\newcommand{\tabref}{Table~\ref}
\newcommand{\neqref}{Eq.~\eqref}
\newcommand{\secref}{Section~\ref}

\def\l{{\bm l}}
\def\bpi{{\bm \pi}}

\newlength{\defaulttabcolsep}
\title{ADVANCING CTC-CRF BASED END-TO-END SPEECH RECOGNITION WITH WORDPIECES AND CONFORMERS}
\name{Huahuan Zheng$^{1}$\sthanks{Equal contribution and random listing.}, Wenjie Peng$^{2}$\footnotemark[1], Zhijian Ou$^{1}$\sthanks{Corresponding author.} and Jinsong Zhang$^{2}$}
\address{
 $^{1}$Speech Processing and Machine Intelligence (SPMI) Lab, Tsinghua University, China\\
 $^{2}$School of Information Science, Beijing Language and Culture University, China \\
}
\begin{document}
\ninept
\maketitle
\begin{abstract}
    Automatic speech recognition systems have been largely improved in the past few decades and current systems are mainly hybrid-based and end-to-end-based. The recently proposed CTC-CRF framework inherits the data-efficiency of the hybrid approach and the simplicity of the end-to-end approach. In this paper, we further advance CTC-CRF based ASR technique with explorations on modeling units and neural architectures. Specifically, we investigate techniques to enable the recently developed wordpiece modeling units and Conformer neural networks to be succesfully applied in CTC-CRFs. Experiments are conducted on two English datasets (Switchboard, Librispeech) and a German dataset from CommonVoice. Experimental results suggest that (i) Conformer can improve the recognition performance significantly; (ii) Wordpiece-based systems perform slightly worse compared with phone-based systems for the target language with a low degree of grapheme-phoneme correspondence (e.g. English), while the two systems can perform equally strong when such degree of correspondence is high for the target language (e.g. German).
\end{abstract}
\section{Introduction}
\label{sec:intro}
A typical Automatic Speech Recognition (ASR) system usually consists of acoustic model (AM), language model (LM) and pronunciation model (PM). 
Depending on how to organize the above components during training and inference, current ASR systems can be divided into two main categories, namely hybrid and end-to-end (E2E) systems. Hybrid systems usually optimize each component separately and combine AM and LM with a PM during inference stage. Such modular optimization of hybrid systems can make full use of data (i.e. unpaired-text for LM training), whereas training (i.e. forced alignment, tree-based clustering, etc.) is very complicated. In contrast, E2E systems aim to simplify the training pipeline and fold all the components into a single neural network (NN) to optimize jointly. There are various types of E2E systems, including E2E-LF-MMI-based \cite{hadian2018end}, CTC-based \cite{graves2006connectionist, amodei2016deep}, Attention-based Seq2Seq \cite{chorowski2014end, chorowski2015attention} and RNN-T \cite{graves2012sequence, rao2017exploring}. Generally, E2E systems can be trained in an elegant way with very promising results but tend to be data-hungry, and require tons of paired speech data. Compared with hybrid and E2E systems, the recently developed CTC-CRF \cite{xiang2019crf, an2020cat} framework inherits the data-efficiency of the hybrid approach and the simplicity of the end-to-end approach.
It has been shown that CTC-CRF outperforms regular CTC consistently on a wide range of benchmarks, and is on par with other state-of-the-art end-to-end models \cite{xiang2019crf,an2020cat,zheng2021efficient}. This work aims to further advance the CTC-CRF based ASR technique with explorations on modeling units and neural architectures.

Recently, wordpiece-based modeling units \cite{rao2017exploring,zhang2020faster} and Conformer neural networks \cite{gulati2020conformer} have demonstrated their effectiveness in other ASR architectures (hybrid, RNN-T). 
It has been shown in \cite{xiang2019crf} that mono-char CTC-CRFs work well, but yield worse results compared with monophone CTC-CRFs. Previous work in CTC-CRFs only use BLSTM \cite{xiang2019crf} and VGG-BLSTM \cite{an2020cat} for neural architectures, which are somewhat old-fashioned.
It is highly possible that wordpiece modeling units and Conformer neural networks can bring performance gains when coupled with the CTC-CRF loss. This paper tries to answer these questions through technique investigation and extensive experiments.

After an overview of related work on wordpiece modeling units and Conformer neural architectures in \secref{sec:relatedwork:unit} and \secref{sec:relatedwork:nn} respectively, we introduce the CTC-CRF framework (\secref{sec:ctc-crf-asr}) and present the techniques pertaining to modeling units, neural architectures, data augmentation and the learning rate scheduler to train Conformers in \secref{sec:train}, which enable the successful application of wordpieces and Conformer in CTC-CRFs. Then, we evaluate the techniques on three datasets (Switchboard, Librispeech and CommonVoice German) respectively in \secref{sec:exp}. The ablation study in \secref{sec:exp:ablation-study} aims to clarify how different techniques affect the recognition performance compared to that of \cite{xiang2019crf, an2020cat}. Our main contribution is that we successfully advance CTC-CRF based ASR techniques with wordpieces and Conformers, and the experimental results are very competitive among the state-of-the-art results on the three datastes.

\section{Related Work}
\label{sec:relatedwork}

\subsection{Modeling unit}
\label{sec:relatedwork:unit}

A basic problem for acoustic modeling is the choice of modeling units. Conventionally, ASR systems (i.e. GMM-HMM \cite{young1995the}, DNN-HMM \cite{hinton2012deep} etc.) heavily rely upon phonetic knowledge. These systems usually adopt phone as modeling units, where AM labels come from word-level transcripts using an expert-curated pronunciation lexicon. For languages like English with a low degree of grapheme-phoneme correspondence, it has been shown that a highly optimized pronunciation lexicon could achieve fairly well results so far \cite{han2021multistream, an2021deformable,zhou2021phoneme}. Despite the promising results of phone-based ASR systems, creating expert-curated pronunciation lexicon is time-consuming. Thus, another direction for AM modeling units is to replace phone with grapheme to eliminate the need of lexicon construction. Early explorations on grapheme-based modeling units have shown that these systems usually perform worse than phone-based systems \cite{kanthak2002context}, while grapheme-based systems could achieve results as well as phone-based systems for languages with high degree of grapheme-phoneme correspondence \cite{killer2003grapheme}. Grapheme-based systems are not better than phone-based systems in English, until taking the context into account \cite{le2019from}. Recently, grapheme-based units like BPE \cite{sennrich2016neural} or wordpiece \cite{wu2016google}, which originate from the task of Neural Machine Translation (NMT), have been broadly applied to ASR systems, including the hybrid \cite{zhang2020faster} and E2E \cite{rao2017exploring} systems.



\subsection{Neural architecture}
\label{sec:relatedwork:nn}

Another problem for acoustic modeling is the appropriate design of AM architecture. Previously, GMM-HMM \cite{young1995the} is the dominant architecture for ASR, where GMM models the emission probabilities while HMM models the transition probabilities. It has been shown that replacing GMM with deep neural networks (DNN) can boost the recognition accuracy \cite {hinton2012deep}. Later on, neural networks have been the dominant architecture for AM and various neural networks have been advanced and proposed. Since speech is a signal with quasi-stationary property, one of the main trends of exploring neural networks for acoustic modeling is to capture the temporal information. For example, RNN and its variants \cite{graves2013speech} have shown to model such information effectively. 

\begin{figure*}[ht]
    \centering
    \includegraphics[width=1.0\linewidth]{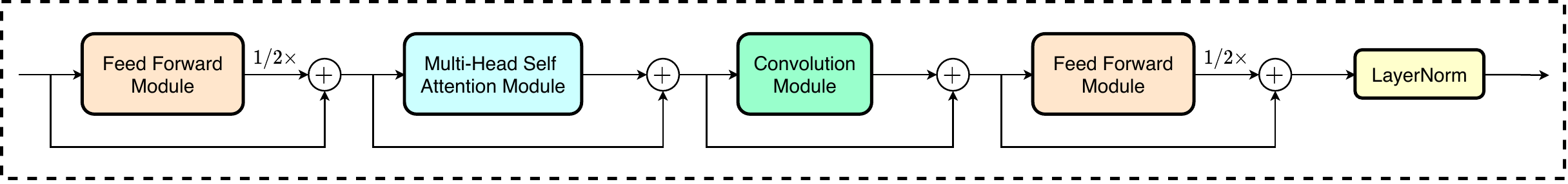}
    \caption{Conformer block architecture. A Conformer block comprises of two macaron-like feed-forward modules with half-step residual connection sandwiching a multi-head self attention module and a convolution module and the following layer normalization.}
    \label{fig:conformer}
\end{figure*}

Recently, the Transformer architecture based on self-attention \cite{karita2019a,wang2020transformer,zhang2020transformer} has been broadly applied to acoustic modeling due to its excellent ability to capture long-term dependency and high training efficiency. In addition to modeling global information, using convolutional receptive fields to capture local information has also been successfully applied in ASR \cite{abdel-hamid2014convolutional,li2019jasper}. To leverage the global and local information simultaneously, \cite{gulati2020conformer} propose Conformer to combine self-attention with convolution, and such combination is illustrated in \figref{fig:conformer}. Since then, Conformer has been successfully applied to several speech processing tasks \cite{guo2021recent}.


\section{CTC-CRF based ASR}
\label{sec:ctc-crf-asr}

In this section, we give a brief review of CTC-CRF based ASR. Basically, CTC-CRF is a conditional random field (CRF) with CTC topology. We first introduce the CTC method. Given an observation sequence $ \bm{x} = (x_{1}, x_{2}, \cdots x_{T})$, i.e. the speech feature sequence, we denote the corresponding label sequence as $\bm{l} = (l_{1}, l_{2}, \cdots, l_{U})$. The CTC loss function is defined as:
\begin{equation}
    \label{eq:obj}
    \mathcal{L}(\theta) = -\text{log}\, p_{\theta}(\bm{l}|\bm{x})
\end{equation}
where $\theta$ is the model parameters. Since $\bm{x}$ and $\bm{l}$ usually differ in length (i.e. $ T \geq U $ ) and are not aligned
in speech recognition, a framewise state sequence $\bm{\pi} = (\pi _{1}, \pi _{2}, \cdots, \pi _{T})$ is introduced in CTC to handle the alignment. In CTC, the probability of a state sequence $\bm{\pi}$ given an observation sequence $\bm{x}$ is defined as: 
%
%
%
\begin{equation}
    \label{eq:ctcpathequation}
    p_{\theta}(\bm{\pi}|\bm{x}) = \prod^{T}_{t=1}p_\theta(\pi_t|\bm{x})
\end{equation}
where $p_{\theta}(\pi _{t} | \bm{x})$ is the label posterior probability at time $t$ given input sequence $\bm{x}$. 
The posterior of $\l$ is defined through the posterior of $\bpi$ as follows:
\begin{equation}
    \label{eq:ctclabelequation}
    p_{\theta}(\bm{l}|\bm{x}) = \sum_{\bm{\pi} \in \mathcal{B}^{-1}(\bm{l})}p_{\theta}(\bm{\pi} | \bm{x})
\end{equation}
where $\mathcal{B}$ is a function, mapping state sequences into label sequences by removing consecutive repetitive labels and blanks \cite{graves2006connectionist}. 
Notably, \neqref{eq:ctcpathequation} assumes that the states between all time steps are conditionally independent. To overcome such unreasonable assumption, CTC-CRF extends CTC and redefine the posterior of $\bpi$ as a CRF:
\begin{equation}
    \label{eq:crfllh}
    p_\theta(\bm{\pi}|\bm{x})=\frac{\exp(\phi_\theta(\bm{\pi}, \bm{x}))}{\sum_{\bm{\pi'}} \exp(\phi_\theta(\bm{\pi'}, \bm{x}))}
\end{equation}
Here $\phi_{\theta}(\bm{\pi}, \bm{x})$ denotes the potential function of the CRF, defined as:
\begin{equation}
    \label{eq:crfpotential}
    \phi_\theta(\bm{\pi}, \bm{x})=\sum_{t=1}^T \log p_\theta (\pi_t|\bm{x}) + \log p(\bm{l})
\end{equation}
where $\bm{l} = \mathcal{B}(\bpi)$. The sum $\sum_{t=1}^T \log p_\theta (\pi_t|\bm{x})$ defines the node potential, calculated from the bottom DNN. $\log p(\bm{l})$ defines the edge potential, realized by an n-gram LM of labels which is often referred to as the denominator n-gram LM. 
Combining Eq. (\ref{eq:ctclabelequation})(\ref{eq:crfllh})(\ref{eq:crfpotential}) yields the CTC-CRF loss function:
\begin{equation}
    \label{eq:ctc-crf-loss}
    \begin{aligned}
        \mathcal{L}(\theta) & = -\log \frac
        {\sum_{\bm{\pi} \in \mathcal{B}^{-1}(\bm{l})} \exp(\phi_\theta(\bm{\pi}, \bm{x}))}
        {\sum_{\bm{\pi'}} \exp(\phi_\theta(\bm{\pi'}, \bm{x}))}
    \end{aligned}
\end{equation}

By incorporating $\log p(\bm{l})$ into the potential function in CTC-CRF, the conditional independence drawback suffered by CTC is naturally avoided. It has been shown that CTC-CRF outperforms regular CTC consistently on a wide range of benchmarks, and is on par with other state-of-the-art end-to-end models \cite{xiang2019crf,an2020cat,zheng2021efficient,an2021deformable}.


\section{Training recipe} 
\label{sec:train}

In this section, we discuss the techniques applied in Conformer AM training, where the whole training pipeline is illustrated in \figref{fig:flowchart}. It can be seen from \figref{fig:flowchart} that such pipeline consists of processing steps from two main perspectives:
\begin{figure}[hbt]
    \centering
    \includegraphics[width=0.48\textwidth]{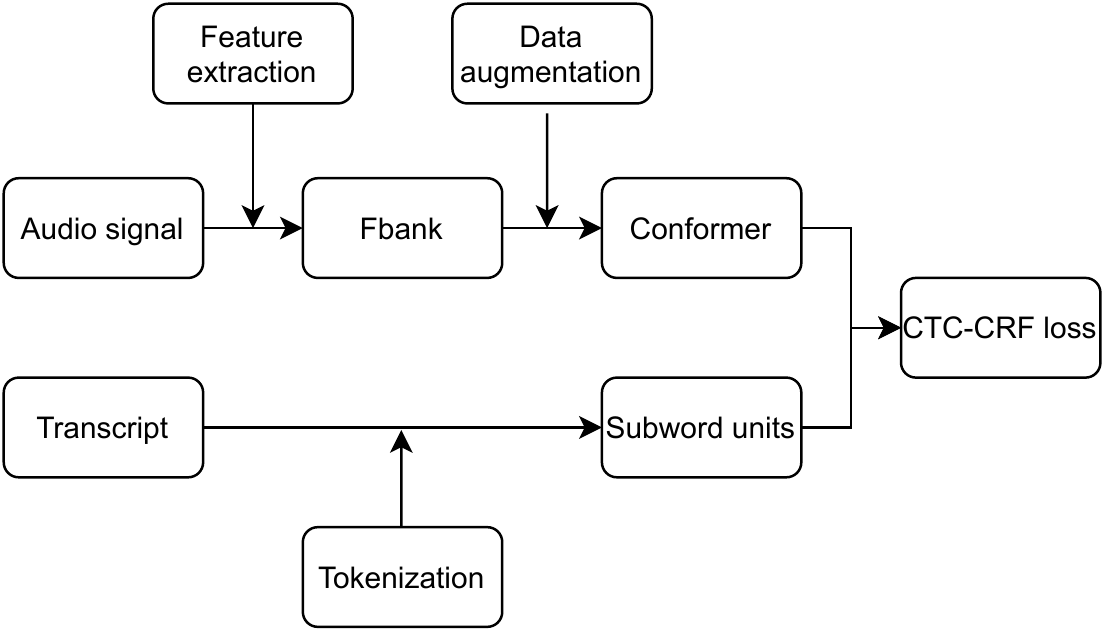}
    \vspace{-6mm}
    \caption{Overall AM training pipeline.}
    \label{fig:flowchart}
\end{figure}

\paragraph*{Data preparation} process is very similar to that of \cite{an2020cat}, including the preparation of input features and subword tokenization for AM respectively. However, this work differs from \cite{an2020cat} at how we prepare these two parts of data. For input features, we extract 80-dimensional Fbank features from a 25ms window with stride of 10ms and normalize the features with cepstral mean and variance normalization (CMVN). Unless otherwise stated, 3-way speed perturbation and \cite{ko2015audio} and SpecAug \cite{park2019specaugment} are adopted as data augmentation. We will present how the new input features influence recognition performance in \secref{sec:exp:ablation-study}. Details of subword tokenization will be present in \secref{sec:train:subword}. We also implement our new SpecAug, which will be introduced in \secref{sec:train:specaug} and analyzed in \secref{sec:exp:ablation-study}.

\paragraph*{AM training} is conducted with Confromer neural networks coupled with the CTC-CRF loss. The speech features are first fed into a 1/4 convolution subsampling layer with or without SpecAug, then a linear layer projects the features to the same dimension as the hidden size of Conformer blocks, and then feeds the features into a series of sequentially stacking Conformer blocks. To better monitor the training of Conformers, we introduce our learning rate scheduler (see \secref{sec:train:scheduler}). Following \cite{gulati2020conformer}, we also examine the influence of Conformer size on the recognition performance in \secref{sec:exp:ablation-study}.

\subsection{Subword tokenization}
\label{sec:train:subword}

For subword tokenization, we implement three systems with modeling units of phones, wordpieces and characters respectively. In the phone-based systems, pronunciation lexicons will be adopted to tokenize word-level transcripts if they are available, otherwise Phonetisaurus Grapheme-to-Phoneme (G2P) \cite{novak2012wfst} will be used. For wordpiece and character based systems, we use SentencePiece \cite{kudo2018sentencepiece} for tokenization. The tokenization processes for character- and wordpiece-based systems are slightly different from the phone-based systems, as the two systems do not generate the actual phonetic lexicons, they just simply define a mapping rule from word to characters or wordpieces. 

For wordpiece-based systems, we first train a tokenization model with the \verb|unigram| mode and the wordpiece size of 150. Wordpieces occured with high frequency in training texts will be reserved while rare ones will be mapped into \verb|<unk>|. We find such mapping is very crucial for AM training. Additionally, \verb|<s>| and \verb|</s>| are excluded since they are not involved in AM training. Consequently, there are 148 wordpieces generated from the trained tokenization model. Then, we utilize the trained tokenization model to encode word-based transcripts into wordpiece ID. For word-to-wordpiece mapping, we tokenize the words occurred in the training set and collect the pairs of word and wordpiece ID. We map words into wordpiece IDs rather than plain wordpiece text, because some characters in the corpus need normalization, otherwise they can not be handled properly. In short, we utilize SentencePiece \cite {kudo2018sentencepiece} to map words into wordpiece IDs for AM training, and revert wordpiece IDs to words using the word-to-wordpieceID mapping rule at the stage of decoding. The data preparation procedure of character-based systems is the same as in wordpiece-based systems. The only difference is that the tokenization mode should be set to \verb|char| when using SentencePiece for tokenization. 


Despite the broad applications of wordpieces as modeling units for ASR systems, there seems to be no consensus on how to set the
appropriate size of the wordpiece set, which is usually determined experimentally \cite{zhang2020faster,rao2017exploring}. 
When the size of the wordpiece set increases, the denominator n-gram LM (see \neqref{eq:crfpotential}) will be enlarged, which will increases the training cost. Thus, we fix the size of the wordpiece set to be 150 throughout all our experiments in this work.

\subsection{SpecAug with ratio}
\label{sec:train:specaug}

SpecAug \cite{park2019specaugment} has shown its power in modern end-to-end ASR systems \cite{gulati2020conformer, saon2021advancing}. The basic idea of SpecAug is to randomly mask some part of the input features along time and frequency dimension respectively and warp the time steps within a given range. The original SpecAug \cite{park2019specaugment} does not take the length of the features into account, while recent works \cite{gulati2020conformer, park2020specaugment} show that when SpecAug considers ratios with respect to the the sequence length in masking and warping, it performs better than the original one. Thus, in this work, we implement our own ratio SpecAug and discuss it in \secref{sec:ablation:specaug}.

\subsection{Learning rate scheduler}
\label{sec:train:scheduler}

The learning rate scheduler of Conformer \cite{gulati2020conformer}, which is borrowed from Transformer training \cite {vaswani2017attention}, includes a linearly warmup and a decreasing decay by the inverse square root to the steps, defined as:
\begin{equation}
    \label{eq:transformerlr}
    lr(n)=d_{model}^{-0.5}\cdot \min(n^{-0.5}, n\cdot N_{warmup}^{-1.5})
\end{equation}
where $n$ and $N_{warmup}$ indicate the number of training steps and warmup steps respectively, and $d_{model}$ indicates the hidden dimension of Conformer blocks. With the Transformer scheduler, training usually terminates when reaching the predefined steps. To add more flexibility to control the peak warmup learning rate, we introduce a factor $p$ to multiply with $d_{model}^{-0.5}$, which is set 0.5 for Librispeech (\secref{sec:exp:libri}) while 1.0 for Switchboard (\secref{sec:exp:swbd}) and CommonVoice German (\secref{sec:exp:cvge}) respectively. We also introduce an early stop mechanism to avoid overfitting, where learning rate will decay with a factor of 0.3 if the loss does not decrease on the validation set. And training will terminate when the learning rate is less than a given threshold.

\section{Experiments}
\label{sec:exp}

We evaluate Conformer AMs with different modeling units on three datasets - the 260-hour Switchboard \cite{godfrey1992switchboard}, the 1000-hour Librispeech \cite{panayotov2015librispeech} and the 700-hour Mozilla CommonVoice 5.1 German (CV German) \cite{ardila2019common}. For Conformer based AM training, we experiment with three configurations in all our experiments and discuss their effects in \secref{sec:ablation:amsize}. To further improve the performance, we also apply Kaldi's latest rescoring scripts \cite{li2021a} with word-level Transformer LMs to rescore the N-best lists generated from the first-pass 4-gram based decoding.

\subsection{Switchboard}
\label{sec:exp:swbd}


The techniques in \secref{sec:train} are applied on the conversational Switchboard dataset. Following \cite {li2021a}, we train Transformer LMs with 6 hidden layers, using the speech transcripts and English Fisher corpus, where there are a total of 34M tokens in the training set. The sizes of vocabularies used for LM training differ slightly between phone-based and wordpiece-based systems, but both are roughly 30K. Such difference originates from the the construction of lexicon as described in \secref{sec:train:subword}. Accordingly, we train Transformer LMs separately for the two systems with the same configuration. The results on Switchboard are shown in \tabref{tab:swbd}.

\renewcommand\tabcolsep{3pt}
\begin{table}[hbt]
    \centering
    \caption{WER results (\%) of different systems on the Eval2000 test set. ``SW'' and ``CH'' are short for Switchboard and Callhome subsets of Eval2000 respectively. ``Trans.'' is short for Transformer neural networks. ``\#Params'' counts the parameters of AM in million. The `+' in the second column indicates doing interpolation of n-gram LMs and NN LMs. The results in square brackets denote the weighted average over SW and CH based on our calculation when not reported in the original paper.}
    \resizebox{0.48\textwidth}{!}{
        \begin{threeparttable}
            \label{tab:swbd}
            \begin{tabular}{llccccc}
                \toprule
                \multicolumn{2}{l}{\textbf{Method}}         & \textbf{Unit}           & \textbf{\#Params}          & \textbf{SW}            & \textbf{CH}  & \textbf{Eval2000}                 \\
                \midrule
                \textbf{RNN-T}                                                                                                                                                                 \\
                BLSTM \cite{saon2021advancing} & \multirow{2}{*}{RNN LM} & \multirow{2}{*}{char}      & \multirow{2}{*}{57}    & 6.6          & 13.9              & 10.3          \\
                \quad + i-vectors                 & ~                       & ~                          & ~                      & 6.4          & 13.4              & 9.9           \\

                \midrule
                \textbf{CTC/Attention}                                                                                                                                                         \\
                Conformer \footnotemark                         & Trans.                  & bpe                     & 44.6                   & 6.8          & 14.0              & 10.4          \\
                \midrule
                \textbf{LF-MMI}                                                                                                                                                                \\
                TDNN-F \cite{li2021a}         & 4-gram                  & \multirow{3}{*}{triphone}     & \multirow{3}{*}{-}     & 8.6          & 17.0              & 12.8          \\
                ~ & +Trans. \tnote{$\ast$}  & ~    & ~ & 7.2          & 14.4              & 10.8          \\
                ~ & +Trans. \tnote{$\ast\ast$}          & ~                          & ~ & 6.5          & 13.9              & 10.2 \\
                \cmidrule{2-7}
                \multirow{2}{*}{TDNN-LSTM \cite{hadian2018flat} }& 4-gram & \multirow{2}{*}{biphone} & \multirow{2}{*}{-} & 9.8 & 19.3 & [14.6] \\
                ~ & +RNN LM & ~ & - & 8.5 & 17.4 & [13.0] \\
                \midrule
                \textbf{CTC-CRF}                                                                                                                                                               \\
                \multirow{2}{*}{VGGBLSTM \cite{an2020cat}}   & 4-gram                  & \multirow{2}{*}{monophone}     & \multirow{2}{*}{39.15} & 9.8          & 18.8              & 14.3          \\
                ~                                           & +RNN LM                  & ~                          & ~                      & 8.8          & 17.4              & 13.1          \\
                \cmidrule{2-7}
                \multirow{4}{*}{\makecell[l]{Conformer\\ (This work) }} & 4-gram & \multirow{2}{*}{monophone} & \multirow{2}{*}{51.82} & 7.9 & 16.1 & 12.1 \\
                ~                                           & +Trans.                  & ~                          & ~                      & 6.9 & 14.5     & 10.7 \\
                \cmidrule{2-7}
                ~                                           & 4-gram                  & \multirow{2}{*}{wordpiece} & \multirow{2}{*}{51.85} & 8.7          & 16.5              & 12.7          \\
                ~                                           & +Trans.                  & ~                          & ~                      & 7.2          & 14.8              & 11.1          \\
                \bottomrule
            \end{tabular}
            \begin{tablenotes}
                \item [$\ast$] N-best rescoring.
                \item [$\ast\ast$] Iterative lattice rescoring.
            \end{tablenotes}

        \end{threeparttable}
    }
\end{table}

\footnotetext{Results of Conformer with 2k bpe from \url{https://github.com/espnet/espnet/tree/master/egs/swbd/asr1}.}

As \tabref{tab:swbd} shows, for monophone CTC-CRFs, using Conformer in this work significantly outperform the prior work of using VGG-BLSTM \cite{an2020cat} with or without NN LM rescoring. Specifically, by using Conformer together with Transformer LM rescoring, this work reduces the WER by 18.32\% (10.7\% vs 13.1\%) against the prior best CTC-CRF system, which uses VGG-BLSTM and RNN LM rescoring in \cite{an2020cat}.
The wordpiece-based CTC-CRF performs very close to monophone CTC-CRF (11.1\% vs 10.7\%), both using Conformers. Considering that English is a language with a low degree of grapheme-phoneme correspondence, this performance gap is not surprising.
Nevertheless, it can be seen that the performance gap between monophone CTC-CTF and wordpiece CTC-CRF becomes smaller, when compared to the gap between monophone CTC-CRF and mono-char CTC-CRF as shown in \cite{xiang2019crf}. This confirms the advantage of using wordpiece units over using characters.
Notably, CTC-CRF is similar to LF-MMI \cite{povey2016purely}, so we also compare the monophone CTC-CRF system with a triphone LF-MMI system as in \cite{li2021a}, and both use the same Transformer LM rescoring scripts. CTC-CRF achieves slightly better results (10.7\% vs 10.8\%) with N-best rescoring, despite LF-MMI is further improved by iterative lattice rescoring, which, however, doubles the computation cost compared to N-best rescoring. We find that for our CTC-CRF based systems, iterative lattice rescoring hardly improve over N-best rescoring. Our systems also exhibit competitive performance when compared to other top performing systems from the literature on the Switchboard dataset.

\subsection{Librispeech}
\label{sec:exp:libri}

The Librispeech dataset includes 960-hour speech data, among which we split 95\% to train the AM and use the rest as the validation set. The official test-clean, test-other, dev-clean and dev-other datasets are for evaluation and excluded from training. In the Librispeech experiments, we only apply our implementation of SpecAug and do not use 3-fold speed perturbation.

We use the openly available 42-layer Transformer LM trained by RWTH \cite{irie2019language}. 
The Transformer LM was trained on the speech transcripts plus an additional 800M-word text-only corpus. The vocabulary size is around 200k. We plug the LM scores calculated from RWTH's LM into Kaldi's Transformer LM rescoring scripts \cite{li2021a} to obtain the N-best rescoring results.
We present the results on Librispeech and compare our results against those from different systems in the literature in \tabref{tab:libri}.

\renewcommand\tabcolsep{3pt}
\begin{table}[ht]
    \centering
    \caption{WER results (\%) of different systems on Librispeech.}
    \resizebox{0.48\textwidth}{!}{
        \begin{threeparttable}
            \label{tab:libri}
            \begin{tabular}{llcccccc}
                \toprule
                \multicolumn{2}{l}{\multirow{2}{*}{\textbf{Method}}}     & \multirow{2}{*}{\textbf{Unit}} & \multirow{2}{*}{\textbf{\#Params}} & \multicolumn{2}{c}{\textbf{test}}                                                    \\
                \cmidrule(l){5-6}
                ~                                                        & ~                              & ~                                  & ~                                & \textbf{clean}                    & \textbf{other} \\
                \midrule
                \multicolumn{2}{l}{\textbf{RNN-T}}                       & ~                              & ~                                  & ~                                & ~                                           \\
                \multicolumn{2}{l}{Conformer \cite{gulati2020conformer}} & wordpiece                      & 118.8                              & 1.9            & 3.9                             \\
                \midrule
                \multicolumn{2}{l}{\textbf{CTC}}                         & ~                              & ~                                  & ~                                & ~                                 \\
                \multirow{2}{*}{\makecell[l] {vggTrans. \cite{zhang2020faster}}}
                                                                         & 4-gram                         & \multirow{2}{*}{wordpiece}         & \multirow{2}{*}{81}              & 2.31           & 4.79           \\
                ~                                                        & + Trans.                    & ~                                  & ~                                & 2.10           & 4.20           \\

                \midrule
                \multicolumn{2}{l}{\textbf{Hybrid}}                      & ~                              & ~                                  & ~                                & ~                                 \\
                \multirow{3}{*}{\makecell[l] {Multistream                                                                                                                                                                                                                \\ CNN \cite{han2021multistream} } }
                                                                         & 4-gram                         & \multirow{3}{*}{triphone}          & \multirow{3}{*}{20}              & 2.80           & 7.06
                \\
                ~                                                        & + RNN LM                 & ~                                  & ~                                & 2.34           & 6.04           \\
                ~                                                        & + SRU LM    & ~                                  & ~                                & 1.75           & 4.46           \\

                \cmidrule{2-6}
                \multirow{2}{*}{\makecell[l]{
                    BLSTM \cite{irie2019language}} } & 4-gram & \multirow{2}{*}{triphone} & - & 3.8 & 8.8 \\
                ~ & +Trans. & ~ & - & 2.5 & 5.7 \\
                \midrule 
                \multicolumn{2}{l}{\textbf{CTC-CRF}}                     & ~                              & ~                                  & ~                                & ~                                 \\
                BLSTM \cite{xiang2019crf} & 4-gram & monophone & 13 & 4.09 & 10.65 \\
                \cmidrule{2-6}
                \multirow{4}{*}{\makecell[l]{
                        Conformer (This work)
                    }}
                                                                         & 4-gram                         & \multirow{2}{*}{monophone}         & \multirow{2}{*}{51.82}           & 3.61           & 8.10           \\
                ~                                                        & + Trans.                    & ~                                  & ~                                & 2.51           & 5.95           \\
                \cmidrule{2-6}
                ~                                                        & 4-gram                         & \multirow{2}{*}{wordpiece}         & \multirow{2}{*}{51.85}           & 3.59           & 8.37           \\
                ~                                                        & + Trans.                    & ~                                  & ~                                & 2.54           & 6.33           \\
                \bottomrule
            \end{tabular}

        \end{threeparttable}
    }
\end{table}

Generally, the main observations from \tabref{tab:libri} over the Librispeech dataset are the same as those from \tabref{tab:swbd} over the Switchboard dataset.
Within the CTC-CRF framework, Conformer neural networks bring significant performance gain, and phone-based and wordpiece-based systems perform equally strong.
Compared to a strong hybrid system \cite{irie2019language} which uses the same Transformer LM, the CTC-CRF systems achieve lower WER before NN LM rescoring, and the WERs are very close after applying NN LM rescoring. 
When compared with a recent top performing system \cite{han2021multistream}, our phone-based systems perform slightly better on test other (5.95\% vs 6.04\%) when applying NN LM rescoring for only one time. When compared with \cite{gulati2020conformer,zhang2020faster}, the current CTC-CRF models are much smaller and could be further improved.


\subsection{Mozilla CommonVoice German}
\label{sec:exp:cvge}
Mozilla CommonVoice is a crowdsourcing, open source dataset including multiple languages. Our experiments are conducted on the CommonVoice 5.1 German dataset, which consists of around 700-hour speech data and paired text. We first conduct the monophone-based experiments with the original data split, but find that the loss over the development set abnormally converges after only 2 epochs of training. Thus, we combine the train and development data and then resplit for training and validation. We apply such data split the same for both character-based and wordpiece-based systems.


The Transformer LMs are trained over the speech transcripts, no extra texts are used, and the hyper-parameters are the same as those in \secref {sec:exp:swbd}. There are about 13M tokens in the training set, and the vocabulary sizes are roughly 157K for all the three systems, based on monophones, characters, and wordpieces respectively. Notably,  wordpiece-based and character-based systems use the same vocabulary, which differs from that in the phone-based system. We train two Transformer LMs separetely with the same configuration to conduct N-best rescoring.

\renewcommand\tabcolsep{\defaulttabcolsep}
\begin{table}[hbt]
    \centering
    \caption{WER results (\%) of different systems on CommonVoice German.}
    \resizebox{0.48\textwidth}{!}{
        \begin{threeparttable}
            \label{tab:cvge}
            \begin{tabular}{llccc}
                \toprule
                \multicolumn{2}{l}{\textbf{Method}} & \textbf{Unit} & \textbf{\#Params}          & \textbf{WER(\%)}              \\
                \midrule
                \multicolumn{2}{l}{\textbf{CTC/Attention}} & ~ & ~ & ~ \\
                Transformer \footnote   & RNN LM & bpe & 27.42 & 10.8 \\
                \midrule
                \multicolumn{2}{l}{\textbf{CTC-CRF}} & ~ & ~ & ~ \\
                \multirow{6}{*}{\makecell[l]{ Conformer (This work)}} & 4-gram & \multirow{2}{*}{char} & \multirow{2}{*}{25.03} & 12.7 \\
                ~                                   & +Trans.        & ~                          & ~                      & 11.6 \\
                \cmidrule{2-5}
                ~                                   & 4-gram        & \multirow{2}{*}{monophone}     & \multirow{2}{*}{25.03} & 10.7 \\
                ~                                   & +Trans.        & ~                          & ~                      & 10.0 \\
                \cmidrule{2-5}
                ~                                   & 4-gram        & \multirow{2}{*}{wordpiece} & \multirow{2}{*}{25.06} & 10.5 \\
                ~                                   & +Trans.        & ~                          & ~                      & 9.8  \\
                \bottomrule
            \end{tabular}

        \end{threeparttable}
        }
\end{table}

\footnotetext{Transformer results from \url{https://github.com/espnet/espnet/tree/master/egs/commonvoice/asr1} }

As shown in \tabref{tab:cvge}, the wordpiece-based CTC-CRF system performs better than the wordpiece-based CTC/Attention system from ESPnet with 9.3\% relative WER reduction.
Further, it can be seen that the performance gap between wordpiece-based and phone-based CTC-CRFs is very close on the CV German dataset. And the wordpiece-based CTC-CRFs even achieve slightly better results compared to phone-based CTC-CRFs. Such observation on the comparison of phone-based and wordpiece-based systems is different from \secref{sec:exp:swbd} and \secref{sec:exp:libri}, where phone-based systems outperform wordpiece-based systems.
An important factor may be the language difference. English has a low degree of grapheme-phoneme correspondence, while such degree of correspondence is high for German. 
Compared to the wordpiece-based systems, the character-based systems perform worse. This again confirms the advantage of using wordpieces over using characters as the units for ASR.


\subsection{Ablation studies}
\label{sec:exp:ablation-study}

Our ablation experiments are conducted on the Switchboard dataset, basically following the settings in \secref {sec:exp:swbd}. Different methods are evaluated under the phone-based systems and the results are from the first-pass 4-gram decoding.

\subsubsection{AM architecture}
\label{sec:ablation:am}

We compare BLSTM, VGGBLSTM with the Conformer architectures in the CTC-CRF systems under the same experimental settings and the same data preparation manner as in \cite{xiang2019crf, an2020cat}. We reduce the hidden size of the Conformer to 128, making the model size comparable to that of the BLSTM in \cite{xiang2019crf}. 
As \tabref{tab:swbd-cmp} shows, compared to BLSTM, Conformer with a similar model-size obtains 5.3\% relative WER reduction. Notably, the Conformer even performs slightly better than the VGGBLSTM that contains more than 3 times of parameters. These results clearly show the advantage of Conformer over BLSTM and VGGBLSTM.

\renewcommand\tabcolsep{3pt}
\begin{table}[ht]
    \centering
    \caption{WER results (\%) of the phone-based systems with different AMs on Switchboard. For fair comparison, 40 Fbank features with $\Delta$ and $\Delta\Delta$ are used as the input features for Conformer with no data augmentation. All results are from 4-gram LM based decoding.}
    \resizebox{0.48\textwidth}{!}{
        \begin{threeparttable}
            \label{tab:swbd-cmp}
            \begin{tabular}{lccccc}
                \toprule
                \textbf{Method}          & \textbf{AM model}         & \textbf{\#Params (M)} & \textbf{SW} & \textbf{CH} & \textbf{Eval2000} \\
                \midrule
                \multirow{3}{*}{CTC-CRF} & BLSTM \cite{xiang2019crf} & 13.47                 & 10.3        & 19.7        & [15]              \\
                ~                        & VGGBLSTM \cite{an2020cat} & 39.15                 & 9.8         & 18.8        & 14.3              \\
                \cmidrule{2-6}
                ~                        & Conformer                 & 12.53                 & 9.2         & 19.2        & 14.2              \\
                \bottomrule
            \end{tabular}

        \end{threeparttable}
    }
\end{table}

\subsubsection{Conformer model size}
\label{sec:ablation:amsize}

We use three Conformer models with 12M, 25M and 51M parameters respectively, denoted by Conformer-S+ (small plus), Conformer-M (medium) and Conformer-M+ (medium plus), which, for clarity, are different from the notations in \cite{gulati2020conformer}. The Conformer-S+ has 16 blocks with hidden size of 180, 4 attention heads and convolution kernel size of 32. As for Conformer-M and Conformer-M+, the corresponding hyper-parameters are (16, 256, 4, 32) and (17, 360, 8, 32). The ablation experiments demonstrate that increasing the model sizes clearly brings performance improvements. 

\renewcommand\tabcolsep{\defaulttabcolsep}
\begin{table}[ht]
    \centering
    \caption{WER results (\%) of the phone-based systems with different Conformer sizes on Switchboard.}
    \begin{threeparttable}
        \label{tab:ablation-size}
        \begin{tabular}{lcccc}
            \toprule
            \textbf{Model} & \textbf{\#Params} & \textbf{SW} & \textbf{CH} & \textbf{Eval2000} \\
            \midrule
            Conformer-S+   & 12.81             & 8.3         & 17.1        & 12.7              \\
            Conformer-M    & 25.03             & 8.2         & 16.4        & 12.3              \\
            Conformer-M+   & 51.82             & 7.9         & 16.1        & 12.1              \\
            \bottomrule
        \end{tabular}
    \end{threeparttable}
\end{table}

\subsubsection{SpecAug}
\label{sec:ablation:specaug}
The effects of SpecAug are analyzed with the Conformer-S+ model. The hyper-parameters of ratio SpecAug are $W=0.2, F=0.15, m_F=2, T=0.05, m_T=2$, whose meanings are the same as those in \cite{park2019specaugment} but $W, F, T$ are denoted in proportions here. All our experiments in this work use the same policy for ratio SpecAug. 
It can be seen from \tabref {tab:ablation-sa} that using SpecAug yields 7.1\% and 9.3\% relative WER reductions on SW and CH respectively, and our ratio SpecAug performs slightly better than the frame-wise-fixed SpecAug in \cite {park2019specaugment}.

\renewcommand\tabcolsep{\defaulttabcolsep}
\begin{table}[ht]
    \centering
    \caption{WER results (\%) for the Conformer-S+ based CTC-CRFs on Switchboard, when without SpecAug, with the SM SpecAug, and with our ratio SpecAug.}
    \begin{threeparttable}
        \label{tab:ablation-sa}
        \begin{tabular}{llccc}
            \toprule
            \textbf{Model}                & \textbf{Diff setting}  & \textbf{SW}  & \textbf{CH}   & \textbf{Eval2000} \\
            \midrule
            \multirow{3}{*}{Conformer-S+} & w/o SpecAug            & 9.1          & 18.9          & 14.0              \\
            ~                             & w/ SM SpecAug\tnote{1} & 8.7          & 17.1          & 13.0              \\
            ~                             & w/ ratio SpecAug       & 8.3 & 17.1 & 12.7     \\
            \bottomrule
        \end{tabular}
        \begin{tablenotes}
            \footnotesize
            \item [1] SM is one of SpecAug hyper-parameter sets for Switchboard dataset in \cite{park2019specaugment}.
        \end{tablenotes}
    \end{threeparttable}
\end{table}

\subsubsection{Input features}
\label{sec:ablation:inputfeat}

Previous works on CTC-CRF \cite{xiang2019crf,an2020cat} use the 120 dimensional input features, which consist of 40 Fbank with delta and delta-delta features, which are fed to BLSTM \cite{xiang2019crf} or VGGBLSTM \cite{an2020cat}.
In this experiment, we follow the same way as VGGBLSTM based CTC-CRF \cite{an2020cat} to take the 120 dimensional input features as 3-channel features instead of flat 120-dimensional ones as used by BLSTM, and feed into the bottom convolution layer. 
It can be seen from \tabref{tab:ablation-feature} that using the delta and delta-delta features has marginal effect on WER results. Increasing the features dimension to 80 is beneficial to the performance.

\renewcommand\tabcolsep{\defaulttabcolsep}
\begin{table}[ht]
    \centering
    \caption{WER results (\%) for the Conformer-S+ based CTC-CRFs on Switchboard, with different input features.}
    \begin{threeparttable}
        \label{tab:ablation-feature}
        \begin{tabular}{llccc}
            \toprule
            \textbf{Model}                & \textbf{Feature}                 & \textbf{SW}  & \textbf{CH}   & \textbf{Eval2000} \\
            \midrule
            \multirow{3}{*}{Conformer-S+} & 40 Fbank+$\Delta$+$\Delta\Delta$ & 9.1          & 17.7          & 13.4              \\
            ~                             & 40 Fbank                         & 8.8          & 17.7          & 13.3              \\
            ~                             & 80 Fbank                         & 8.3 & 17.1 & 12.7     \\
            \bottomrule
        \end{tabular}
    \end{threeparttable}
\end{table}

\section{Conclusion}
\label{sec:conclusion}

In this paper, we successfully advance CTC-CRF based ASR techniques with wordpiece modeling units and Conformer neural networks. Several other techniques, including input features, data augmentation and model size, are also thoroughly examined. We find that (i) Conformer can improve the performance significantly and (ii) wordpiece-based systems perform slightly worse than phone-based systems on the two English datasets (Switchobard and Librispeech), while the two systems perform equally strong on the German dataset (CommonVoice). Notably, English and German are two representative languages which have low and high degrees of grapheme-phoneme correspondence respectively. Our results provide good implication for unit selection in ASR.

Overall, this work demonstrates the potential of the CTC-CRF framework for ASR, which can absorb new neural architectures and achieve state-of-the-art results with or without a pronunciation lexicon.
The code will be available at the open-source CAT toolkit \cite{an2020cat}\footnote{\url{https://github.com/thu-spmi/CAT}} for reproducing the results.

\section{Acknowledgements}
We would like to thank Xiaohui Zhang and Ke Li for helpful discussions.


\bibliographystyle{template/IEEEbib}
\bibliography{references/main}

\end{document}